\documentclass[floats,floatfix,showpacs,amssymb,prd,twocolumn,superscriptaddress,nofootinbib,nolongbibliography,reprint]{revtex4-2}

\usepackage{amssymb,amsmath,verbatim,mathtools,needspace,enumitem,etoolbox,graphicx,physics,microtype,afterpage,bigints,gensymb,tabularx,mathtools}

\usepackage[dvipsnames, usenames]{xcolor}
\definecolor{linkcolor}{rgb}{0.0,0.3,0.5}
\definecolor{dodgerblue}{HTML}{1E90FF}
\usepackage[unicode, colorlinks=true, linkcolor=linkcolor, citecolor=linkcolor, filecolor=linkcolor,urlcolor=linkcolor, pdfusetitle]{hyperref}
\usepackage[all]{hypcap}
\usepackage[T1]{fontenc}
\usepackage[utf8]{inputenc}
\usepackage{orcidlink}
\usepackage{tabularx}
\usepackage{bigints}

\makeatletter

\usepackage{scalerel}
\usepackage{stackengine,wasysym}

\newcommand{\ssim}{\mathchar"5218\relax\,}

\makeatletter
\newcommand*{\balancecolsandclearpage}{\close@column@grid \cleardoublepage \twocolumngrid}
\makeatother

\usepackage{lipsum}

\def\apj{{Astrophys.~J.}}\def\apjl{{Astrophys.~J.~Lett.}}
\def\apjs{{Astrophys.~J.~Supp.}}

\def\mnras{{Mon.~Not.~R.~Astron.~Soc.}}

\def\prd{{Phys.~Rev.~D}}

\def\prx{{Phys.~Rev.~X}}
\def\prl{{Phys.~Rev.~Lett.}}

\def\physrep{{Phys.~Rep.}}

\def\lrr{{Living Rev. Relativ.}}

\def\natastro{{Nat. Astron.}}

\newcommand{\bham}{\affiliation{School of Physics and Astronomy \& Institute for Gravitational Wave Astronomy, University of Birmingham, \\ Birmingham, B15 2TT, United Kingdom}}

\newcommand{\milan}{\affiliation{Dipartimento di Fisica ``G. Occhialini'', Universit\'a degli Studi di Milano-Bicocca, Piazza della Scienza 3, 20126 Milano, Italy}}
\newcommand{\infn}{\affiliation{INFN, Sezione di Milano-Bicocca, Piazza della Scienza 3, 20126 Milano, Italy}}

\definecolor{lightgray}{gray}{0.9}

\begin{document}
\title{
Parameter estimation of binary black holes in the endpoint of the up--down instability
}

\author{Viola De Renzis$\,$\orcidlink{0000-0001-7038-735X}}
\email{v.derenzis@campus.unimib.it}
\milan \infn

\author{Davide Gerosa$\,$\orcidlink{0000-0002-0933-3579}}

\milan \infn \bham

 \author{Matthew Mould$\,$\orcidlink{0000-0001-5460-2910}}
  \bham

 \author{Riccardo Buscicchio$\,$\orcidlink{0000-0002-7387-6754}}
\milan \infn

 \author{Lorenzo Zanga$\,$\orcidlink{0009-0004-6981-7008}}

\milan


\date{\today}

\begin{abstract}
Black-hole binary spin precession admits equilibrium solutions corresponding to systems with (anti-) aligned spins. Among these, binaries in the up--down configuration, where the spin of the heavier (lighter) black hole is co- (counter-) aligned with the orbital angular momentum, might be unstable to small perturbations of the spin directions. The occurrence of the up--down instability  leads to gravitational-wave sources that formed with aligned spins but are detected with precessing spins. We present a Bayesian procedure based on the Savage-Dickey density ratio to test the up--down origin of gravitational-wave events. This is applied to both simulated signals, which indicate that achieving strong evidence is within the reach of current experiments, and the LIGO/Virgo events released to date, which indicate that current data are not informative enough.

\end{abstract}

\maketitle


\section{Introduction}

Gravitational-wave (GW) detections provide measurements of the intrinsic properties of astrophysical black holes (BHs), notably their masses and spins. At the time of writing, ground-based interferometers LIGO and Virgo have observed about $ 70$ 
mergers of stellar-mass BHs with false alarm rates $< 1~\mathrm{yr}^{-1}$~\cite{2019PhRvX...9c1040A,2021PhRvX..11b1053A,2021arXiv210801045T,2021arXiv211103606T}
and substantially more detections are expected from the upcoming observing runs
~\cite{2019PhRvD.100f4060B, 2020LRR....23....3A}.

GWs emitted during the inspiral of BH binaries are mostly beamed along the direction of the orbital angular momentum $\boldsymbol{L}$. 
If the spins of the two BHs $\boldsymbol{S}_{1,2}$ are misaligned with  $\boldsymbol{L}$, couplings between these three momenta cause them to precess~\cite{1994PhRvD..49.6274A,1995PhRvD..52..821K}. 
The resulting motion imparts characteristic modulations to the amplitude and phase of emitted GWs. From an astrophysical perspective, measuring spin precession is important to elucidate the possible astrophysical formation pathways of BH binaries, with large spin misalignments thought to be indicative of sources formed via dynamical interactions~\cite{2022PhR...955....1M,2021hgwa.bookE..16M}. 

Configurations with spins that are either aligned or anti-aligned with the orbital angular momentum are equilibrium solutions of the relativistic spin-precession equations.  This means that binaries that are \emph{exactly} aligned will remain so. There are four such cases, which we refer to as up--up, down--down, down--up, and up--down, where ``up'' (``down'') indicates spins that are parallel (anti-parallel) to the orbital angular momentum and the direction before (after) the hyphen refers to the more (less) massive BH. 
Crucially, equilibrium does not imply stability. Reference~\cite{2015PhRvL.115n1102G} showed that, while up--up, down--down, and down--up binaries are always stable, 
up--down binaries 
can be unstable to spin precession. For these sources, infinitesimal perturbations to the spin directions cause large precession cycles. In particular, up--down binaries are stable at early times and turn unstable at the critical orbital separation~\cite{2015PhRvL.115n1102G}
\begin{equation}
r_\mathrm{UD+} = \frac{\left(\sqrt{\chi_1}+\sqrt{q\chi_2}\right)^4}{(1-q)^2}M\,,
\label{eq:rplus}
\end{equation}
where $\chi_{i}=S_{i}/m_{i}^2$ are the Kerr parameters of the BHs, $q=m_2/m_1\leq 1$ is the mass ratio, and $M=m_1+m_2$ is the total mass of the system.\footnote{Throughout the paper we use natural units where $c=G=1$.} The up--down instability was first derived using a Post-Newtonian (PN) approach \cite{2015PhRvL.115n1102G} and then confirmed using both independent PN codes~\cite{2016PhRvD..93l4074L,2022PhRvD.106b3001J} and numerical-relativity simulations \cite{2021PhRvD.103f4003V}.

Measuring the up--down instability in GW data would provide a direct observation of an exquisite feature of the two-body problem in general relativity. At the same time, the up--down instability might also dilute the effectiveness of the spin orientations in discriminating BH-binary formation channels: GW sources that are observed with precessing spins in the LIGO/Virgo band did not necessarily form with misaligned spins. Rather, the spins used to be (anti-) aligned and became misaligned before merger. The flip side of the same coin is that observing unstable binaries will point toward a formation channel that can conceivably explain binaries with up--down spins. Notably, this might include AGN disks surrounding supermassive BHs~\cite{2020MNRAS.494.1203M,2021NatAs...5..749G}, where the spins of embedded stellar-mass BH binaries are expected to either align or anti-align with the angular momentum of the disk~\cite{1975ApJ...195L..65B}.

The up--down instability provides a testable prediction for GW
observations.
Reference \cite{2020PhRvD.101l4037M} showed that unstable up--down BHs do not disperse in the available parameter space but converge to a well-defined endpoint late in the inspiral. This is a precessing configuration where all three angular momenta $\boldsymbol{S}_{1}, \boldsymbol{S}_{2}$, and $\boldsymbol{L}$ are coplanar, and furthermore, the two BH spins are
collinear,
namely \cite{2020PhRvD.101l4037M}, 
\begin{align}
\cos\theta_1&=\frac{\chi_1-q\chi_2}{\chi_1+q\chi_2}\,, \label{eq:tilt1} \\
\cos\theta_2&=\frac{\chi_1-q\chi_2}{\chi_1+q\chi_2}\,, \label{eq:tilt2}\\
\phi_{12}&=0\,,\label{eq:phi12}
\end{align}
where $\theta_{i}$ indicate the tilts angles between $\boldsymbol{S}_i$ and $\boldsymbol{L}$, and  $\phi_{12}$ indicates the azimuthal angle between the two BH spins measured in the orbital plane. After the instability is triggered, binaries reach this analytical endpoint
after the orbital separation has decreased by only
$\lesssim100M$~\cite{2020PhRvD.101l4037M}. Therefore, binaries that form as up--down
and become unstable will appear in our detectors with spin orientations that are well approximated  by Eqs.~(\ref{eq:tilt1}--\ref{eq:phi12}).

In this paper, we perform Bayesian parameter estimation of precessing BH binaries in the endpoint of the up--down instability. Should an unstable up--down binary enter the LIGO band,
can we tell that this source was originally stable and aligned? In statistical terms, this is a model-selection problem between a broader hypothesis where binaries are generically precessing and a narrower hypothesis with constraints given by Eqs.~(\ref{eq:tilt1}--\ref{eq:phi12}). We apply this line of reasoning to both
simulated signals 
and
the current catalog of GW events.
By employing the Savage-Dickey density ratio, we compute the odds in favor of the up--down hypothesis over that of generically precessing BH binaries. Crucially, this only requires an inference run with the uninformative prior, with the odds computed by post-processing the recovered posterior samples.

In Sec.~\ref{sec:Methods} we derive the statistical framework and describe how it can be used to assess whether observed binaries are in the endpoint of the up--down instability. In Sec.~\ref{sec:results} we present our results for an injection campaign and real sources, and also demonstrate that evolving binary BH spin posteriors backwards in time is a useful diagnostic when investigating the up--down instability. We finish with our conclusions in Sec.~\ref{sec:conclusions}.




\section{Methods} 
\label{sec:Methods}


\subsection{Gravitational-wave signals}
\label{subsec:PE_vanilla}
We first consider synthetic GW signals from individual binary BH coalescences on quasi-circular orbits and target the statistical inference of all 15 parameters of the problem. These are two detector frame masses $m_{1,2}$, six spin degrees of freedom (magnitudes $\chi_{1,2}$, tilts $\theta_{1,2}$, azimuthal angles $\phi_{12}$ and $\phi_{JL}$), and seven extrinsic parameters (luminosity distance $D_{L}$, sky location $\alpha,\delta$, polar angle $\theta_{JN}$, polarization $\psi$, coalescence time $t_c$, and phase $\phi_c$). 

 
Signals are analyzed using the parallel version of the {\sc bilby} inference 
code \cite{2019ApJS..241...27A,2020MNRAS.498.4492S}. We use the {\sc IMRPhenomXPHM} approximant \cite{2021PhRvD.103j4056P} for both injection and recovery. 
We consider a three-detector network made of LIGO Livingston, LIGO Hanford, and Virgo at the sensitivity expected for the upcoming O4 run. We use data segments of $4\,\mathrm{s}$, a sampling frequency of $2048\,\mathrm{Hz}$, a low-frequency cutoff of $20\,\mathrm{Hz}$, and zero noise. Spin orientations are quoted at a reference frequency of $20\,\mathrm{Hz}$.  We use the {\sc dynesty} sampler \cite{2020MNRAS.493.3132S} with 2048 live points, a random walk sampling method, a number of autocorrelation equal to 50, and a likelihood that is marginalized over time and distance.

Our priors are those commonly used in the standard LIGO/Virgo analyses \cite{2019PhRvX...9c1040A, 2021PhRvX..11b1053A, 2021arXiv210801045T, 2021arXiv211103606T}. 
In particular, detector-frame component masses are distributed uniformly in $m_{1,2}\in [5,100] M_{\odot}$ with bounds in mass ratio $q \in [1/8, 1]$ and detector-frame chirp mass $\mathcal{M} \in [10, 60]M_{\odot}$ while spins are distributed uniformly in magnitude $\chi_{1,2}\in [0,0.99]$ and isotropically in directions.


In the following, we also postprocess GW data using publicly available posterior samples for the GWTC-2.1~\cite{2021arXiv210801045T} 
and the GWTC-3~\cite{2021arXiv211103606T} 
data releases. Among the available datasets, we use results from the {\sc IMRPhenomXPHM} waveform model where the merger rate is uniform in comoving volume and source-frame time. We consider binary BH mergers with false alarm rates $< 1~\mathrm{yr}^{-1}$ in at least one of the detection pipelines. From these, we exclude all the events that potentially contain a neutron star. 
The resulting list of 69 events is reported in Table~\ref{tab:GWevents}.

When needed, we covert between PN orbital separation $r$ and GW frequency $f_{\rm ref}$ using the 2PN expressions  from Ref.~\cite{1995PhRvD..52..821K}.


%
%
%
%
%
%
%

\renewcommand{\arraystretch}{1.1}
\begin{table}

{\centering
\begin{tabular}{c|c @{\hspace{1em}} || @{\hspace{1em}} c|c}
Event &  $\ln\mathcal{B}$ & Event & $\ln\mathcal{B}$ \\
	\hline\hline
GW150914 &  $0.14$ & GW190731\_140936 &   $0.11$\\
GW151012 & $0.54$ & GW190803\_022701 & $0.11$\\
GW151226 &  $0.50$ & GW190805\_211137 &  $0.61$ \\
GW170104 & $-0.02$ & GW190828\_063405 & $0.3$ \\
GW170608 &  $0.18$ &GW190828\_065509 &  $0.15$\\
GW170729 & $0.47$ & GW190910\_112807 & $-0.06$ \\
GW170809 &  $0.26$ & GW190915\_235702 &  $0.29$ \\
GW170814 &  $-0.06$ & GW190924\_021846 & $0.31$  \\
GW170818 &  $0.58$ & GW190925\_232845 & $0.24$ \\
GW170823 & $0.26$ & GW190929\_012149 &  $-0.15$ \\
GW190408\_181802 & $0.02$ & GW190930\_133541 & $0.59$ \\
GW190412 & $0.6$  & GW191103\_012549 &  $0.58$ \\
GW190413\_052954 &  $-0.01$ & GW191105\_143521 & $0.06$  \\
GW190413\_134308 &  $0.07$ & GW191109\_010717 &$-0.83$ \\
GW190421\_213856 &  $0.09$ & GW191127\_050227 &  $0.31$ \\
GW190503\_185404 &  $-0.04$ & GW191129\_134029 &  $0.33$ \\
GW190512\_180714 &  $0.33$ & GW191204\_171526 & $0.79$ \\
GW190513\_205428 & $0.48$ & GW191215\_223052 &  $0.11$ \\

GW190514\_065416 & $-0.01$ & GW191216\_213338 & $0.27$  \\
GW190517\_055101 & $0.53$ & GW191222\_033537 & $-0.16$ \\
GW190519\_153544 &  $0.35$ & GW191230\_180458 & $0.25$ \\
GW190521 & $-0.26$ & GW200112\_155838 & $0.07$\\
GW190521\_074359 &  $-0.42$ & GW200128\_022011 & $0.46$\\
GW190527\_092055 &  $0.23$ & GW200129\_065458 &  $0.63$\\
GW190602\_175927 & $0.44$ & GW200202\_154313 &  $0.1$ \\
GW190620\_030421 & $0.52$ & GW200208\_130117 & $-0.04$ \\
GW190630\_185205 & $-0.15$ & GW200209\_085452 &  $0.21$\\
GW190701\_203306 &  $0.05$ & GW200216\_220804 &  $0.26$ \\
GW190706\_222641 & $0.8$ & GW200219\_094415 & $0.05$\\
GW190707\_093326 &   $0.04$& GW200224\_222234 & $0.2$\\
GW190708\_232457 &  $0.15$  & GW200225\_060421 & $-0.11$ \\
GW190720\_000836 &  $0.58$ & GW200302\_015811 & $0.05$ \\
GW190725\_174728 &   $0.39$& GW200311\_115853 & $0.32$ \\
GW190727\_060333 &  $0.44$ & GW200316\_215756 & $0.57$\\
GW190728\_064510 &  $0.32$ & &
\end{tabular}
}

\caption{Current GW events and their Bayes factors in favor of the up--down hypothesis
over generic spin precession.
We select events with false alarm rates $< 1~\mathrm{yr}^{-1}$ in at least one of the LIGO/Virgo searches, excluding those that can potentially include a neutron star.
}
\label{tab:GWevents}
\end{table}
\renewcommand{\arraystretch}{1}

\subsection{Savage-Dickey density ratio}
\label{subsec:modelselection}
Given the data $d$ associated with a measurement, and model hypothesis $\mathcal{H}$ characterized by parameters $\theta$, the Bayesian evidence is defined as 
\begin{equation}
\mathcal{Z}(d | \mathcal{H})=\int  \mathcal{L}(d | \theta, \mathcal{H})\pi(\theta | \mathcal{H}) \,\dd\theta \,,
\label{eq:evidence}
\end{equation}
where $\mathcal{L}$ is the likelihood and $\pi$ is the prior distribution. 
Model selection in favor of, say, a ``narrow'' model $\mathcal{H}_{{\rm N}}$ over a ``broad'' model $\mathcal{H}_{{\rm B}}$ 
requires computing the posterior odds 
\begin{equation}
\mathcal{O}=\frac{\mathcal{Z}(d | \mathcal{H}_{{\rm N}}) }{ \mathcal{Z}(d |\mathcal{H}_{{\rm B}} )}
\frac{\pi( \mathcal{H}_{{\rm N}})}{\pi( \mathcal{H}_{{\rm B}})} \,, 
\label{eq:bayesfactor}
\end{equation}
where the first term (ratio of the evidences) is the Bayes factor $\mathcal{B}$. 
Values of the posterior odds are often associated to descriptive terms using the so-called Jeffrey scale~\cite{1963PhT....16c..68J}, where $ | \ln \mathcal{O}|<1$ is classified as ``inconclusive,'' $1< |\ln \mathcal{O}| <2.5$ is classified as as ``weak'' evidence, $2.5<|\ln \mathcal{O}|<5$ is classified as ``moderate'' evidence, and $|\ln \mathcal{O}|>5$ is classified as ``strong'' evidence.  The sign of the log Bayes factor indicates which of the two models is statistically favored, with $\ln\mathcal{O}>0$ signaling a preference for $\mathcal{H}_{{\rm N}}$ over $\mathcal{H}_{{\rm B}}$. 
In the following, we consider equal model priors such that $\mathcal{O} = \mathcal{B}$. 

Let us now assume that model $\mathcal{H}_{{\rm N}}$ is nested within $\mathcal{H}_{{\rm B}}$. That is, among the parameters $\theta=\{\varphi,\gamma\}$, a subset of parameters $\varphi$ is common to both models, while the other parameters $\gamma$ are constrained to  $\gamma_{\rm N}(\varphi)$  in the narrow model. Let us also assume that the prior on $\varphi$ is the same for the two models. In symbols, this is 
\begin{equation}
\pi(\varphi | \mathcal{H}_{{\rm N}}) = \pi(\varphi| \gamma=\gamma_{\rm N}(\varphi), \mathcal{H}_{{\rm B}} )\,.
\end{equation}
Within these assumptions, the Bayes factor in favor of the narrow model reduces to 
\begin{equation}
\mathcal{B}=\bigintsss  \frac{\displaystyle  p(\varphi, \gamma=\gamma_{\rm N}(\varphi) | d, \mathcal{H}_{\rm B}) }{\displaystyle \int  \pi(\varphi',\gamma=\gamma_{\rm N}(\varphi) | \mathcal{H}_{\rm B}) \dd\varphi'} \; \dd\varphi
\,.
\label{eq:BF_SDR0}
\end{equation}
A formal proof of Eq.~(\ref{eq:BF_SDR0}) is presented in Appendix~\ref{savdic}. For the specific case where $\gamma_N$ does not depend on $\varphi$, one has
\begin{equation}
\mathcal{B}=  \frac{p(\gamma=\gamma_{\rm N} | d, \mathcal{H}_{\rm B}) }{  \pi(\gamma=\gamma_{\rm N} | \mathcal{H}_{\rm B})} \;  \,,
\label{eq:BF_SDR}
\end{equation}
where the numerator (denominator) corresponds to the posterior (prior) marginalized over the common parameters $\varphi$. Equation (\ref{eq:BF_SDR}) is the so-called Savage-Dickey density ratio~\cite{2013PLoSO...859655P}.  The key, practical advantage of both these expressions is that they only depend on the broad model $\mathcal{H}_{\rm B}$. One does \emph{not} need to perform inference in the narrow model $\mathcal{H}_{\rm N}$, which can be challenging for non-trivial {submanifolds} $\gamma_{\rm N}(\varphi)$.
It is sufficient to sample the broad model $\mathcal{H}_{\rm B}$ and then evaluate the resulting posterior and prior {probability densities} at the location prescribed by the narrow model. 

\subsection{Application to up--down binaries}
\label{appupdown}
 
For the specific case we are addressing here, the broad model $\mathcal{H}_{\rm B}$ is that of generically precessing BH binaries described in  Sec.~\ref{subsec:PE_vanilla}.
The narrow model $\mathcal{H}_{\rm N}$ consists of binaries in the endpoint of the up--down instability, which are subject to the three constraints of Eqs.~(\ref{eq:tilt1}--\ref{eq:phi12}). 
%
%
%
%
%
From these, we define the parameters $\gamma=\{\gamma_1,\gamma_2,\gamma_3\}$, where
\begin{align}
\label{delta1}
\gamma_1&=\frac{\cos\theta_1 - \cos\theta_\mathrm{UD}(q,\chi_1,\chi_2)}{2}\,, \\
\gamma_2&=\frac{\cos\theta_2 - \cos\theta_\mathrm{UD}(q,\chi_1,\chi_2)}{2}\,, \\
\gamma_3&=\frac{1}{\pi}\arctan\bigg(\dfrac{\sin\phi_{12}}{\cos\phi_{12}}\bigg)\,,
\label{delta3}
\end{align}
and
\begin{align}
\cos\theta_\mathrm{UD}(q,\chi_1,\chi_2) = \frac{\chi_1-q\chi_2}{\chi_1+q\chi_2}\,.
\label{costhetaupdown}
\end{align}
While not unique, we find this parametrization convenient because all the $\gamma_i$ are defined\footnote{The trigonometric manipulation in Eq.~(\ref{delta3}) is necessary because $\phi_{12}\in[0,2\pi]$.} in $[-1,1]$ and the up--down endpoint is mapped to $\gamma=(0,0,0)$. 
%
%
%
%
%
%
%
%
We apply the transformations of Eqs.~(\ref{delta1}--\ref{delta3}) to both prior and posterior samples, estimate the corresponding probability density functions using  
three-dimensional Kernel Density Estimation (KDE), and evaluate the Bayes factor from Eq.~(\ref{eq:BF_SDR}). We use Gaussian kernels and a bandwidth of 0.2~\cite{2021JOSS....6.2784K}.

\begin{figure}
    \includegraphics[width=\columnwidth]{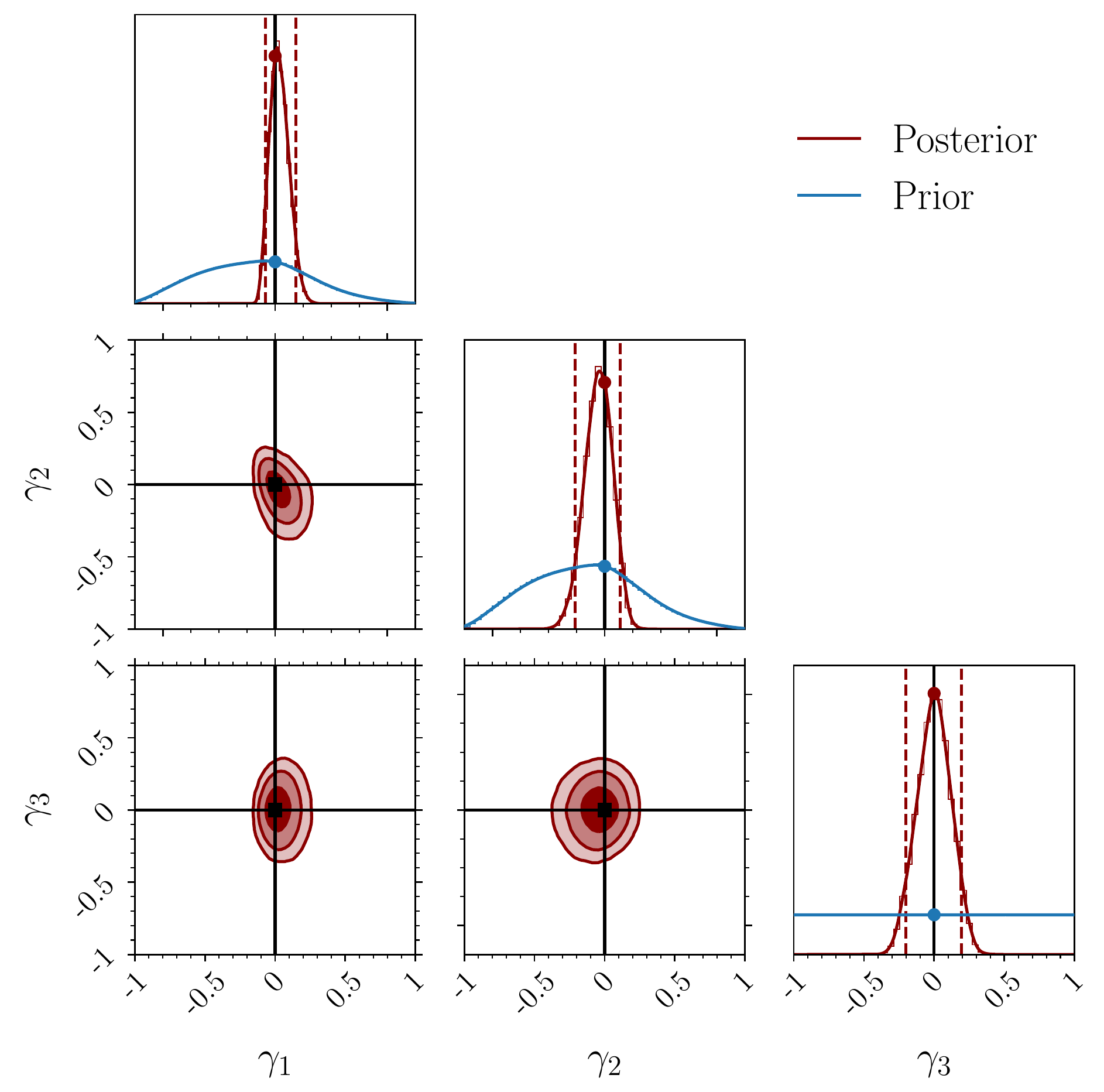}
    \caption{Joint posterior distributions of the rescaled parameters $\gamma = \{\gamma_1,\gamma_2,\gamma_3\}$ defined in Eqs.~(\ref{delta1}--\ref{delta3}). Contour levels correspond to 50\%, 90\%, and 99\% 
    credible regions.  
    Red dashed lines in the 1D marginals indicate the 90\% credible intervals. Solid black lines mark the location of the narrow model $\gamma=0$, i.e., the endpoint of the up--down instability. Black scatter points indicate the value of the posterior (red) and prior (blue) distributions at the endpoint, which are the key ingredients entering the Savage-Dickey evaluation of the Bayesian odds.  
    }   \label{fig:sphere_SDR}
\end{figure}

\begin{figure*}[htb]
    \centering 
  \includegraphics[width=0.325\textwidth]{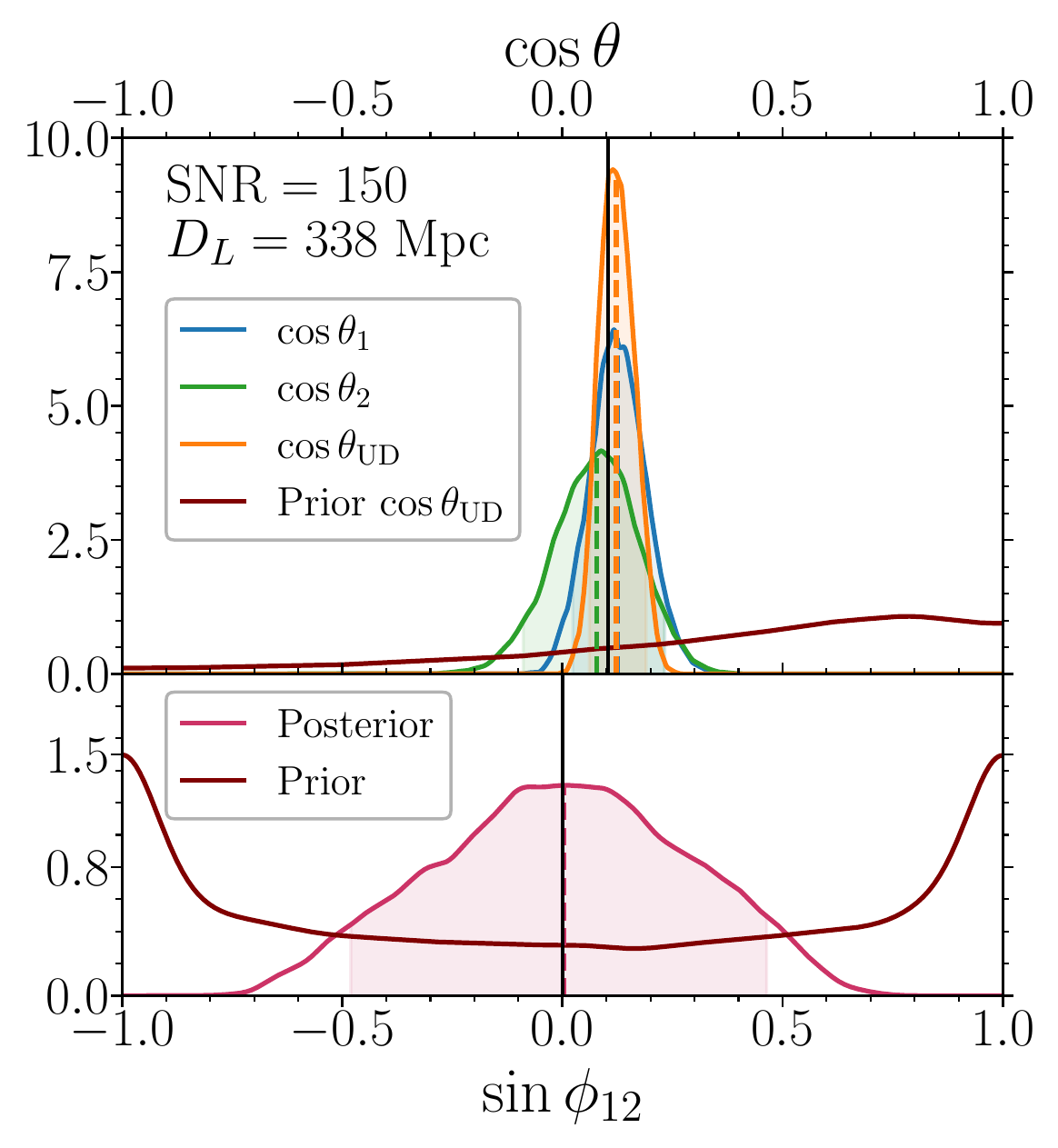}
  \includegraphics[width=0.325\textwidth]{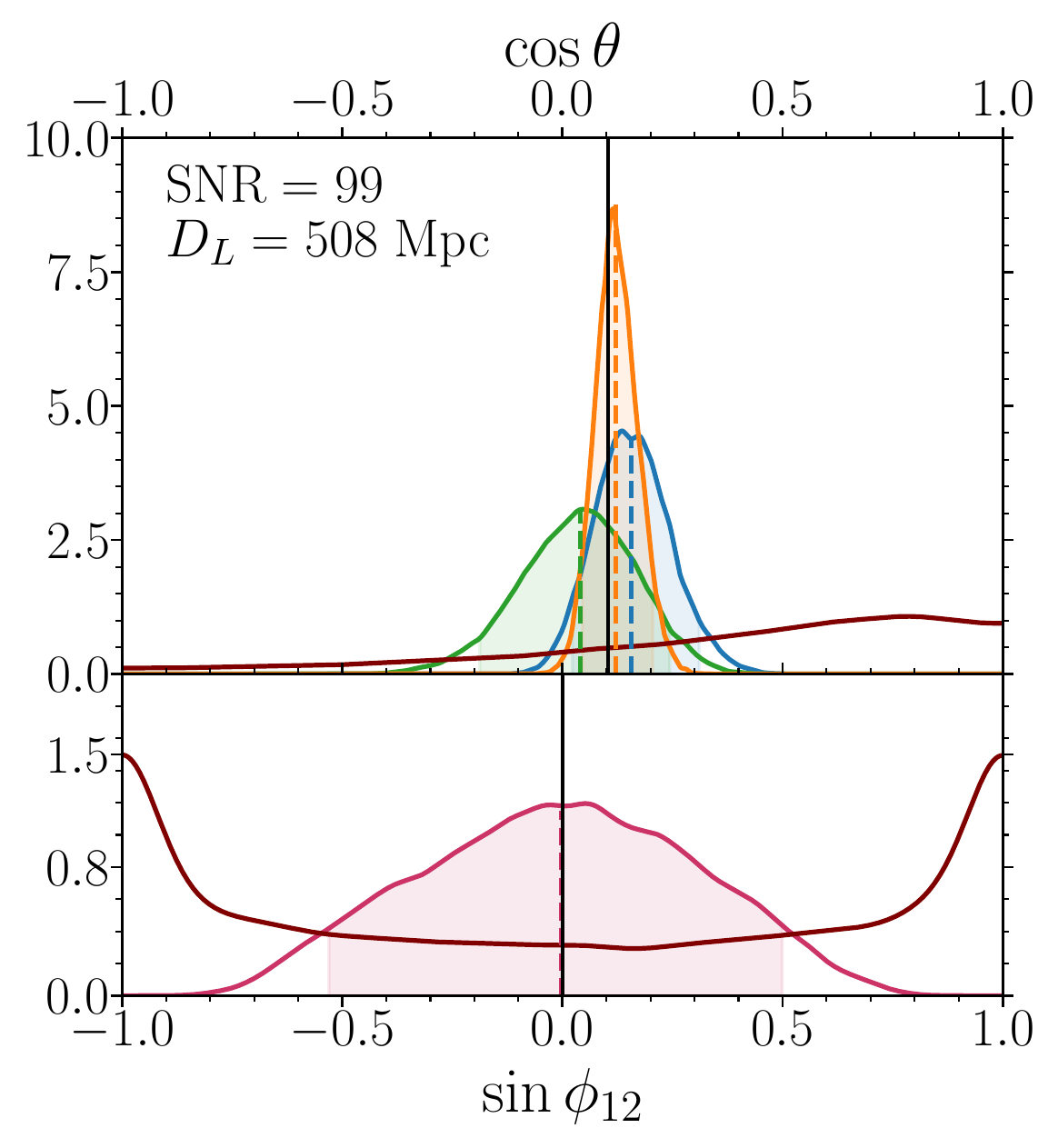}
  \includegraphics[width=0.325\textwidth]{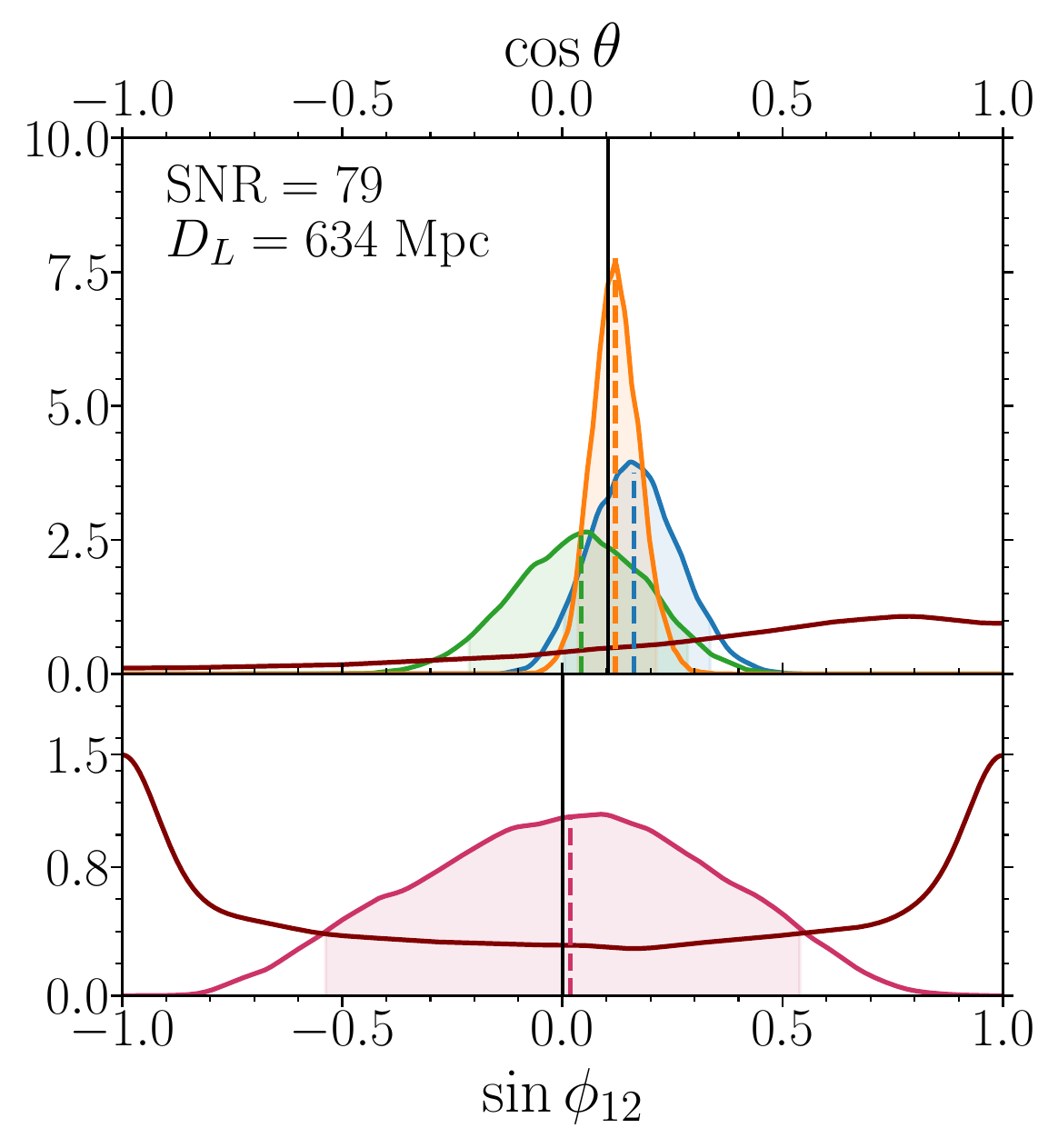}

  \includegraphics[width=0.325\textwidth]{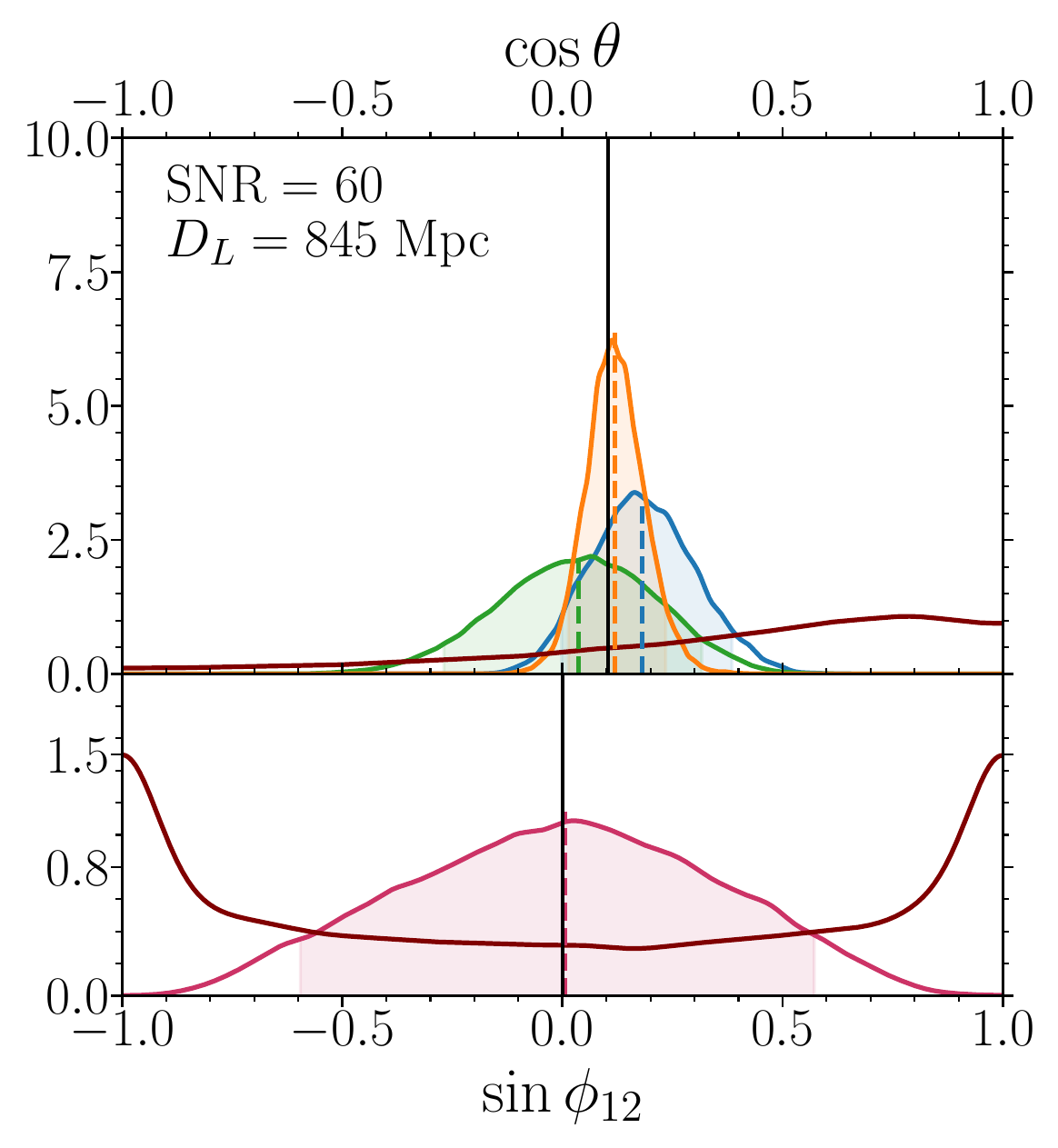}
  \includegraphics[width=0.325\textwidth]{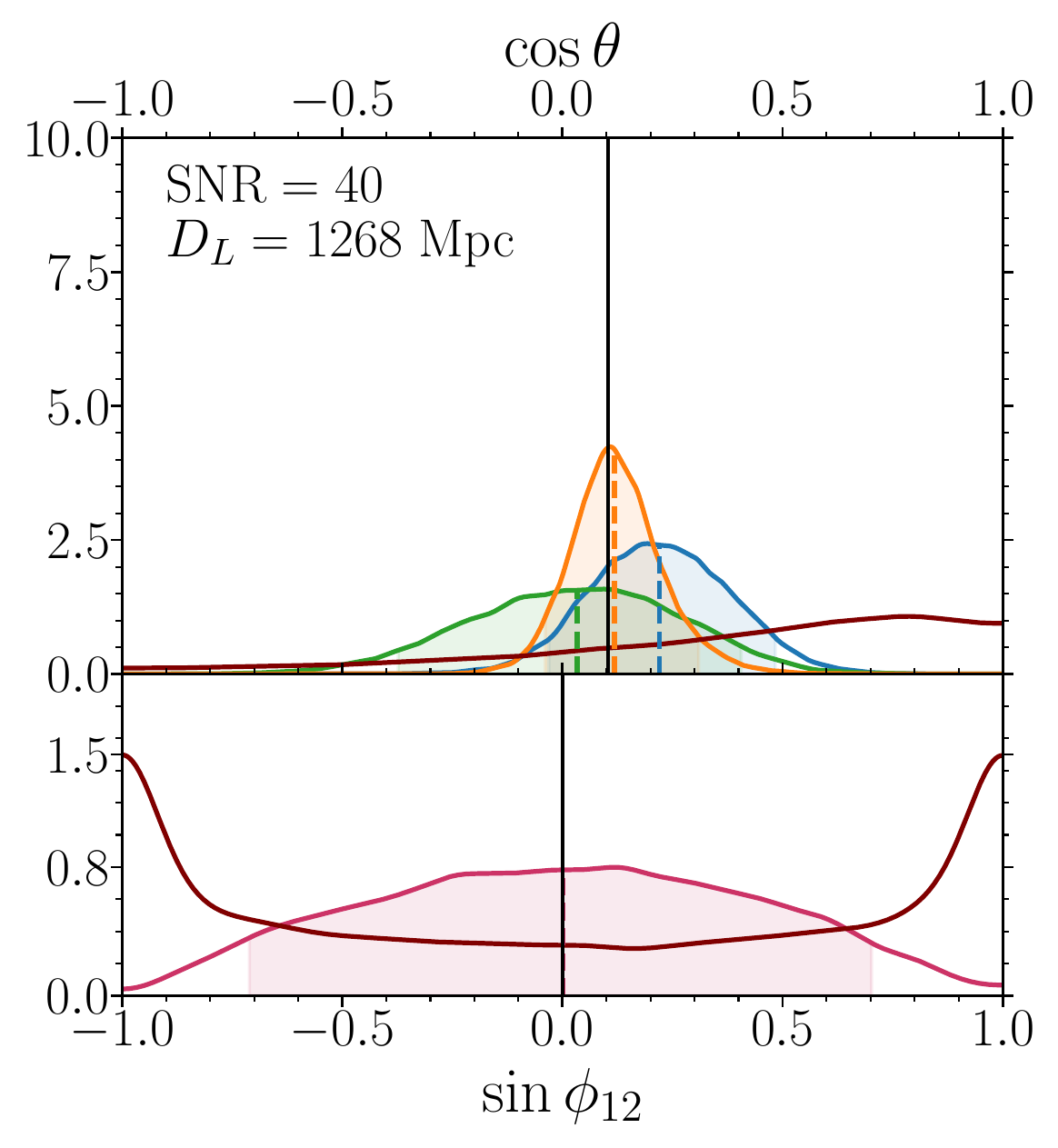}
  \includegraphics[width=0.325\textwidth]{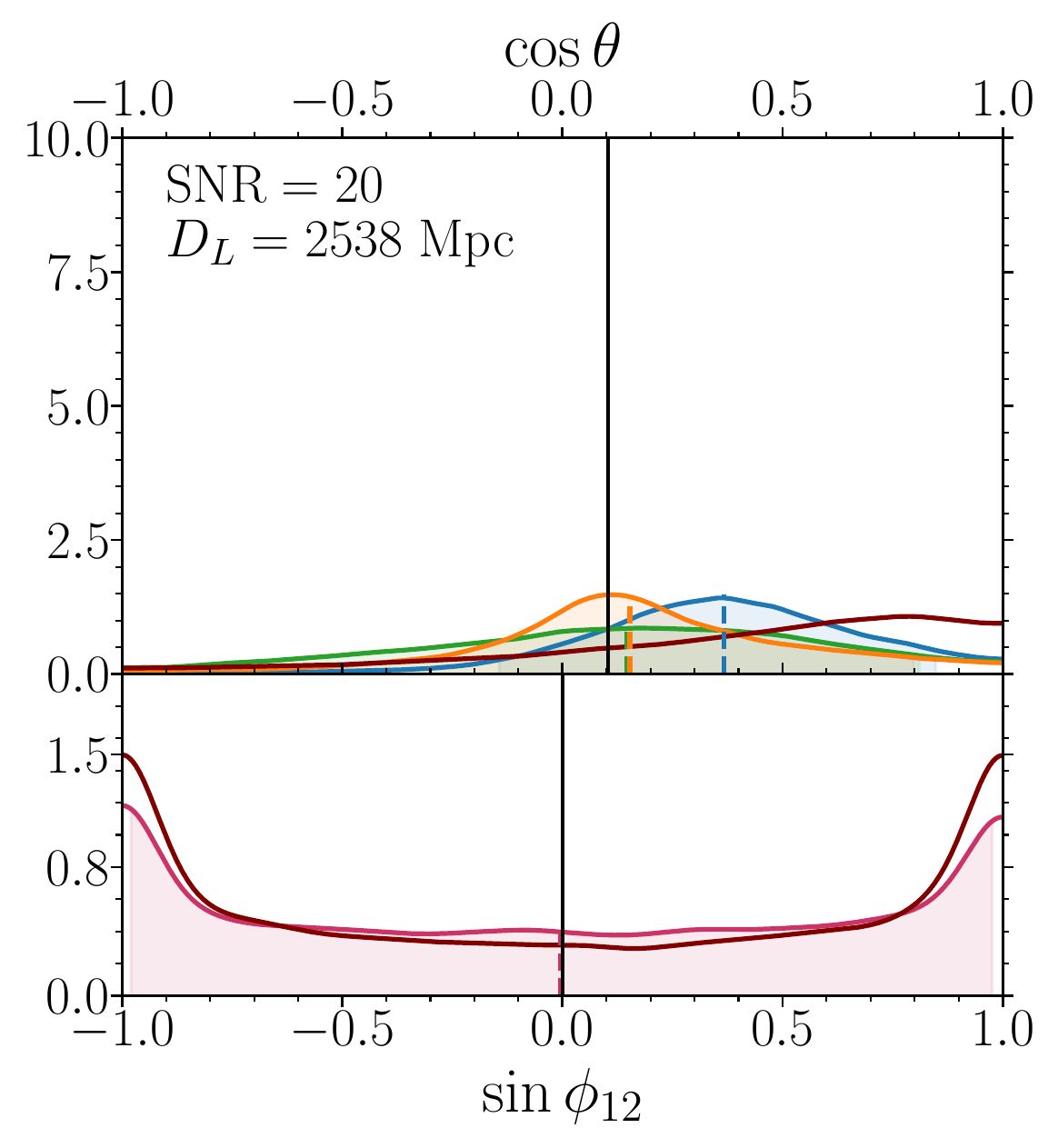}
\caption{Panels from left to right and from top to bottom show parameter-estimation results for the same GW source injected at decreasing (increasing) values of the SNR (luminosity distance $D_{L}$). The upper subpanels show posterior distributions of $\cos\theta_{1}$ (blue), $\cos\theta_2$ (green), $\cos\theta_\mathrm{UD}$ (orange) and the prior distribution of $\cos\theta_\mathrm{UD}$ (dark red); the prior distributions of $\cos\theta_{1,2}$ are flat. The lower subpanels show  posterior (pink) and prior (dark red) distributions of $\sin\phi_{12}$. Black vertical lines indicate the injected values. Dashed vertical lines mark the medians of each distribution while shaded areas indicate the 90\% credible intervals.
}
\label{fig:PosteriorSERIES}
\end{figure*}

An example of this procedure is shown in Fig.~\ref{fig:sphere_SDR}. 
We consider a synthetic source in the endpoint of the up--down instability with tilt angles $\cos\theta_1=\cos\theta_2=\cos\theta_\mathrm{UD}=0.103$ and $\phi_{12}=0$. The injected system has $m_1=49.5M_\odot$  $m_2=39.4M_\odot$, $\chi_1=0.92$, $\chi_2=0.94$, $D_{L}=845$ Mpc, $\theta_{JN}=0.37$, $\phi_{JL}=5.71$, $\alpha=6.11$, $\delta=0.24$, $\psi=2.28$, $t_c=-0.069$ s (in GPS time), and $\phi_c=5.12$.
The prior and posterior KDEs are evaluated at the origin of the $\{\gamma_1,\gamma_2,\gamma_3\}$ cube (black lines in Fig.~\ref{fig:sphere_SDR}). The Savage-Dickey estimate of the Bayes factor is $\ln\mathcal{B} = 5.11$. For equal priors, this corresponds to strong evidence that the source is indeed in the up--down endpoint.
Figure~\ref{fig:sphere_SDR} also shows that the posteriors of the rescaled parameters $\gamma_i$ are somewhat close  to a multivarate Gaussian distribution; this not a generic feature but rather a consequence of the relatively high signal-to-noise ratio (SNR), which for this specific injection is~$60$. 

\section{Results}


\label{sec:results}

\subsection{Comparing posteriors}
\label{subsec:singleinjection}
Before reporting Bayes factors, it is informative to compare posterior distributions against the predictions of Eqs.~(\ref{eq:tilt1}--\ref{eq:phi12}). This a preliminary step which is often used to identify promising candidates for a model-selection analysis. 
  
We consider six synthetic signals describing binary BHs that are in the endpoint of the up--down instability when entering the LIGO band at the reference frequency of $20$ Hz. We use the same set of source parameters as in Fig.~\ref{fig:sphere_SDR}. In particular, we fix the detector-frame masses and inject source waveforms
with $\text{SNR}=150,100,80,60,40,20$, corresponding to  luminosity distances $D_{L}=338,508,634,845,1268, 2538$~Mpc. The PN orbital separation of the binary at $f_{\rm ref}= 20$ Hz is $r_{\rm 20Hz}\simeq 10M$, 
while the critical separation for the instability is $r_{\rm UD+}=266M$. The condition $r_{\rm UD+} - r_{\rm 20Hz} > 100 M$ 
ensures that the predicted endpoint well describes these unstable up--down sources \cite{2020PhRvD.101l4037M}.

Our results are shown in Fig.~\ref{fig:PosteriorSERIES}, where each panel correspond to a different source.  The upper  subpanels compare the posterior distributions of $\cos\theta_{1,2}$ (as obtained from our parameter-estimation analysis) against that of $\cos\theta_\mathrm{UD}$ 
[as obtained from substituting the posterior samples of  $q,\chi_{1},\chi_2$ into Eq.~(\ref{costhetaupdown})]. Note how the prior distribution of  $\cos\theta_\mathrm{ud}$ peaks toward positive values, while those of $\cos\theta_{1,2}$ are flat. Close agreement between the posteriors of $\cos\theta_1$, $\cos\theta_2$, and $\cos\theta_\mathrm{UD}$  provide a qualitative (but not quantitative) indication that the theoretical prediction of the up--down instability is a reasonable description of the data. The lower subpanels report the  posterior distribution of $\sin\phi_{12}$, where values close to zero $0$ indicate a preference for the up--down hypothesis. 

As expected, posteriors for the lowest SNRs tend to cover a large portion of prior range. As the SNR increases, the recovered posteriors approach the injected values that define the endpoint of the up--down instability. In particular, for the case of the highest $\text{SNR}=150$, we find  $\cos\theta_\mathrm{UD}= 0.122_{-0.061}^{+0.068}$ and  $\phi_{12}=0.004_{-0.485}^{+0.460}$ 
(where we quote the median and 90\% credible interval),
compared to the injected values
$\cos\theta_\mathrm{UD}= 0.103$ and $\phi_{12}=0$.

Note that systematic effects are not captured 
in both these results and the rest of the paper because we perform zero-noise runs and use the same waveform model for both injection and recovery.  Waveform systematics in the specific region of parameter space where the up--down instability take place still need to be investigated.

We further note a common feature that characterize all cases shown in Fig.~\ref{fig:PosteriorSERIES}, including those at low SNR. While the recovered values of $\cos\theta_{1,2}$ depart from the injected values as the SNR decreases, the medians of $\cos\theta_\mathrm{UD}$ tend to remain closer to that of the injected endpoint. This seems to indicate that, if the source is truly in the endpoint of the up--down instability, the estimator $\cos\theta_\mathrm{UD}$ might be more accurate than $\cos\theta_{1,2}$. We interpret this as a consequence of more accurate measurements of $q$ and $\chi_{1,2}$ compared those of the spin tilts. This implies we can measure what the endpoint of a binary \emph{would} be from the $q$--$\chi$ posteriors. However, inferring that the given source is in fact in its endpoint requires computation of posterior odds.

\subsection{Model selection}
\label{subsec:backprop}

\begin{figure}
    \includegraphics[width=\columnwidth]{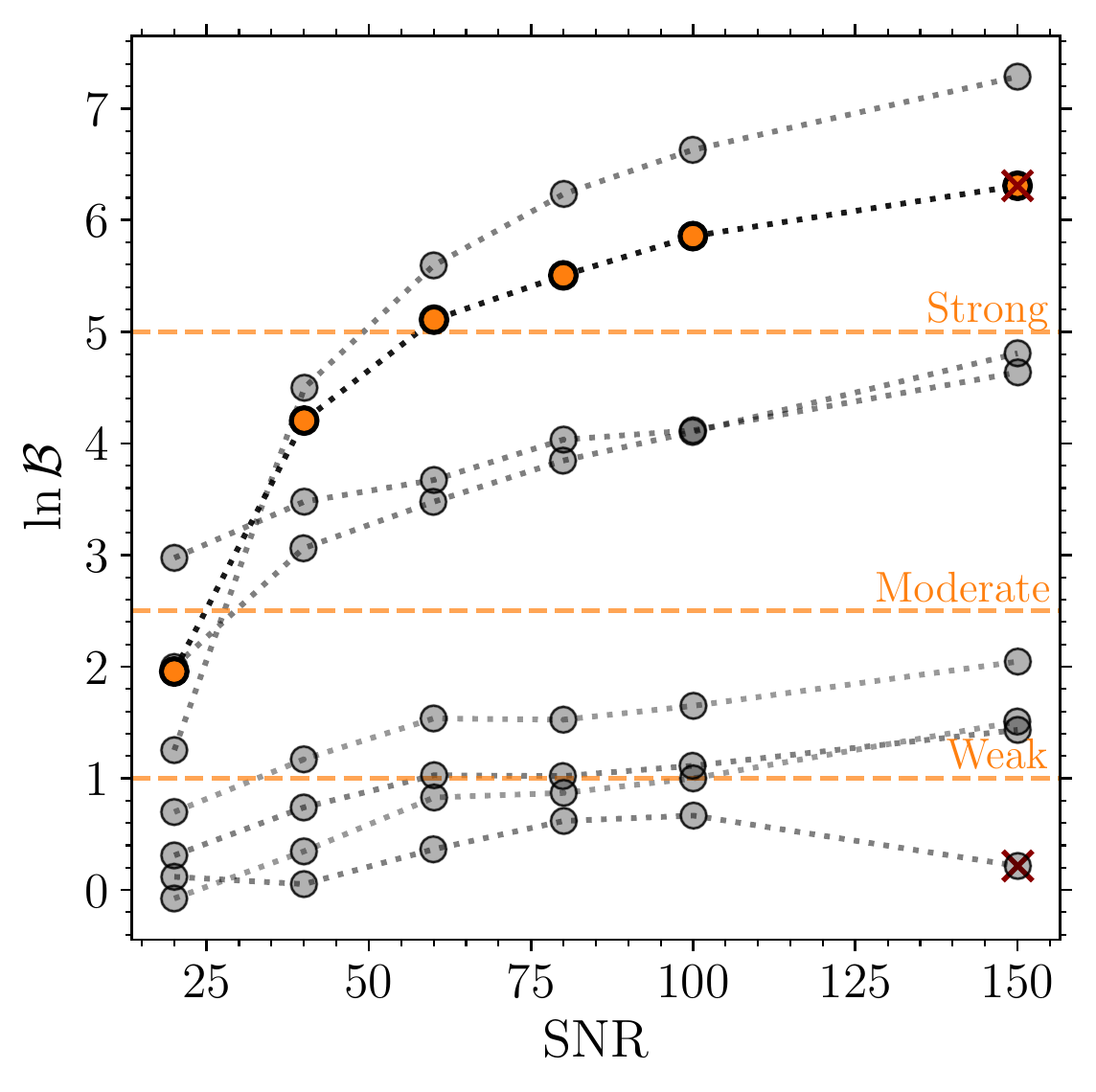}
    \caption{Natural logarithm of the Bayes factor in favor of the up--down hypothesis as a function of the SNR. We consider the same  sources  as in Fig.~\ref{fig:PosteriorSERIES} (orange scatter points) as well as six other  series 
of BH binaries in the up--down endpoint (gray scatter points). Horizontal dashed lines indicate the  threshold values of the Bayes factor in the Jeffrey scale.  Crosses indicate the sources shown in Fig.~\ref{fig:Backprop}. 
}   \label{fig:SeriesSNR}
\end{figure}

While comparing posteriors as in Fig.~\ref{fig:PosteriorSERIES} provides a useful indication of a potential up--down signature, this statement needs to be quantified with a full Bayesian model selection. 
For the same series of six injections, Fig.~\ref{fig:SeriesSNR}  shows the Bayes factor in favor of the up--down hypothesis
over that of generic BH binaries
computed using the Savage-Dickey density ratio (orange points). 
The Bayes factor increases from $\ln \mathcal{B}\ssim 1.96$ for $\mathrm{SNR}=20$ (weak evidence) to $\ln \mathcal{B}\ssim 6.89$ for $\mathrm{SNR}=150$ (strong evidence). 
While this is a controlled experiment where the true source parameters are injected in the up--down configuration, the successful recovery of a large value of $\mathcal{B}$ indicates that data are informative about this property in a concrete measurement setting.

We repeat the same study for six additional series of BH binaries in the up--down endpoint with different parameters $\theta$ 
(gray points) which are part of the broader set of injections described in Sec.~\ref{subsec:100injections}. 
As expected, the Bayes factor increases with the SNR in all cases, though the overall normalization depends on the other source parameters. 
For the case discussed above and shown with orange scatter points, a strong evidence in favor of the nested model is achieved at $\text{SNR}\gtrsim60$ ---values within reach of next LIGO-Virgo observing run~\cite{2020LRR....23....3A}.
However, this is not generic. We find that the distinguishability power critically depends on the source parameters. 
Even among this limited set, there are cases that provide 
only weak or even inconclusive evidence
even at $\mathrm{SNR}=150$. 


\begin{figure*}
    \includegraphics[width=\columnwidth]{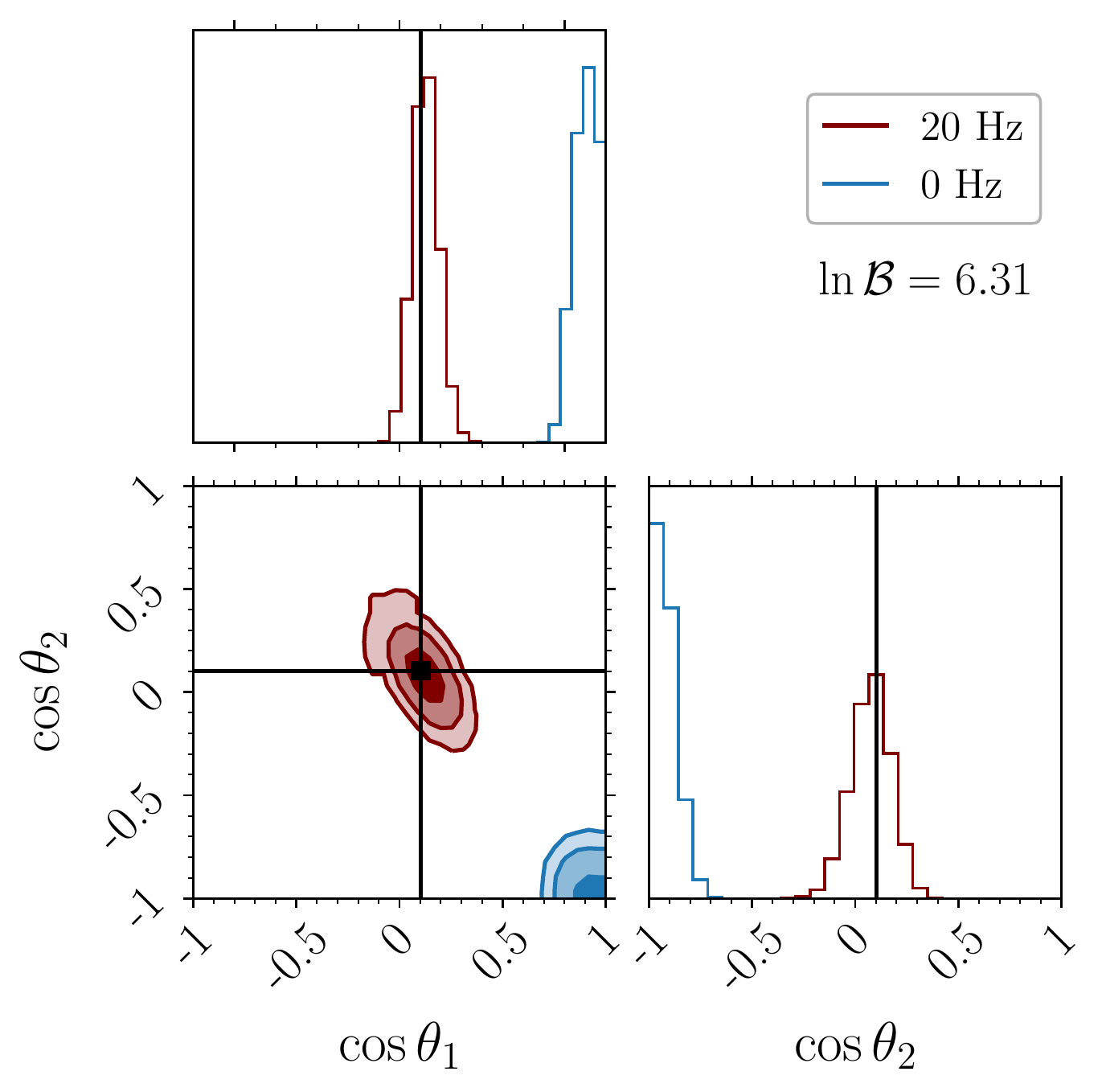}
    \hfill
\includegraphics[width=\columnwidth]{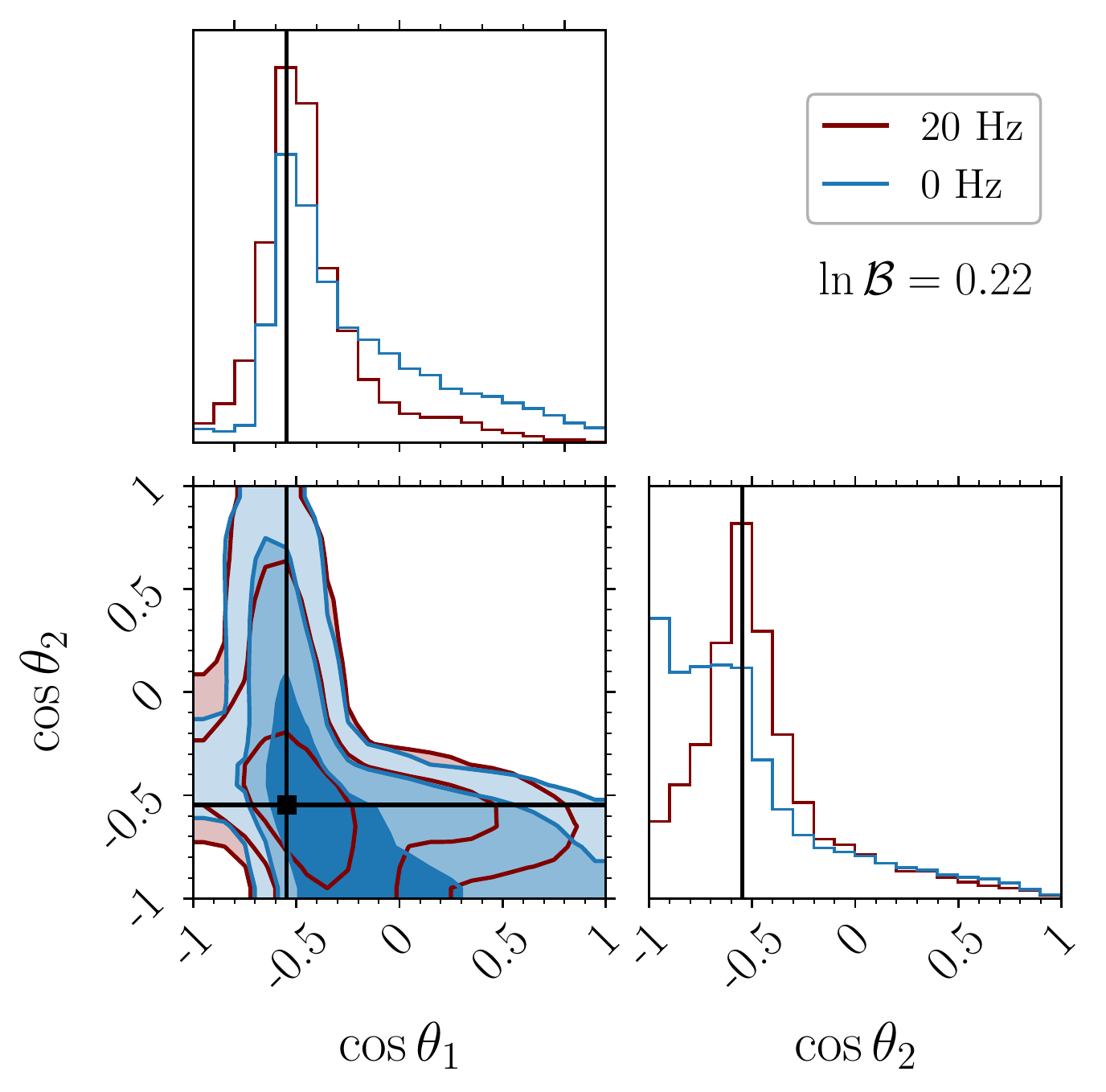}
    
    \caption{Joint posterior distribution of the tilt angles $\theta_1$ and $\theta_2$ for the sources described in Sec.~\ref{backpropsec} and marked with crosses in Fig.~\ref{fig:SeriesSNR}. The left (right) panel shows a case that presents strong (inconclusive) evidence in favor of the up--down hypothesis. Posterior samples are evolved numerically from $f_{\rm ref}= 20$ Hz (red) to $0$ Hz (blue). Solid black lines indicate the injected values. Contour levels mark the 50\%, 90\%, and 99\% credible regions. 
    }
\label{fig:Backprop}
\end{figure*}



\subsection{Backpropagation}
\label{backpropsec}

We can further visualize the up--down signature of BH binaries by back-propagating posteriors samples~\cite{2022PhRvD.105b4076M,2022PhRvD.106b3001J}.
If a detected source is truly an unstable up–down binary, evolving it backward in time should allow us to see it in the up–down spin configuration instead of the particular precessing configuration as observed.
For a given injection, we numerically evolve each posterior sample backward from detection at $f_{\rm ref}=20$~Hz to past-time infinity at $f_{\rm ref}=0$~Hz using precession-averaged PN equations as implemented in Refs.~\cite{2016PhRvD..93l4066G,2023arXiv230404801G}. This procedure requires $q$, $\chi_{1,2}$, $\theta_{1,2}$, $\phi_{12}$,
and $r$ at $f_{\rm ref} = 20$~Hz as inputs and returns the values of the tilt angles $\theta_{12}$ at $0$~Hz ($\phi_{12}$ does not enter the dynamics at infinitely large orbital separations~\cite{2015PhRvD..92f4016G, 2023arXiv230404801G}).

Figure~\ref{fig:Backprop} shows two examples which were selected  from those of Fig.~\ref{fig:SeriesSNR}. Both sources have $\mathrm{SNR}=150$; one provides strong evidence in favor of the up--down endpoint (left panel, $\ln\mathcal{B}=6.31$) while the other returns an inconclusive result (right panel, $\ln\mathcal{B}=0.22$). The parameters of the former are listed in Sec.~\ref{appupdown}  while those of the latter are
$m_1=26M_\odot$, $m_2=26M_\odot$, $\chi_1=0.17$, $\chi_2=0.57$, $\theta_{12}=2.15$, $\phi_{12}=0$, $D_L=190.06$~Mpc, $\psi=2.89$, $\phi=3.33$, $\alpha=3.78$, $\delta=-0.081$, $\theta_{JN}=0.41$, $\phi_{JL}=3.71$, and $t_c=-0.01$ s.



For the binary with large $\mathcal{B}$ (left panel in Fig.~\ref{fig:Backprop}), the posterior distribution at $0$ Hz  is constrained to be close to an aligned binary with up--down spins. 
 In particular, we find $\cos\theta_1>0.80$
 and $\cos\theta_2<-0.99$ at $90\%$
confidence.
This result is an additional, visual indication that data taken at $\ssim20$ Hz are well described by a BH binary that \emph{used to be} aligned but is being observed precessing. 

On the other hand, for the inconclusive case (right panel in Fig.~\ref{fig:Backprop}), the joint distribution of $\cos\theta_1$ and $\cos\theta_2$ at $f_{\rm ref}=0$ Hz occupies a much broader region of the prior volume ($\cos\theta_1>-0.65$ and $\cos\theta_2<0.41$ at $90\%$
confidence).
As indicated by the Bayes factor, this is a source where data are  compatible with a variety of precessing configurations, some that did and some that did not form with up--down spin directions.   


%

\subsection{Injection campaign}
\label{subsec:100injections}

\begin{figure*}
\centering
    \includegraphics[width=0.65\textwidth]{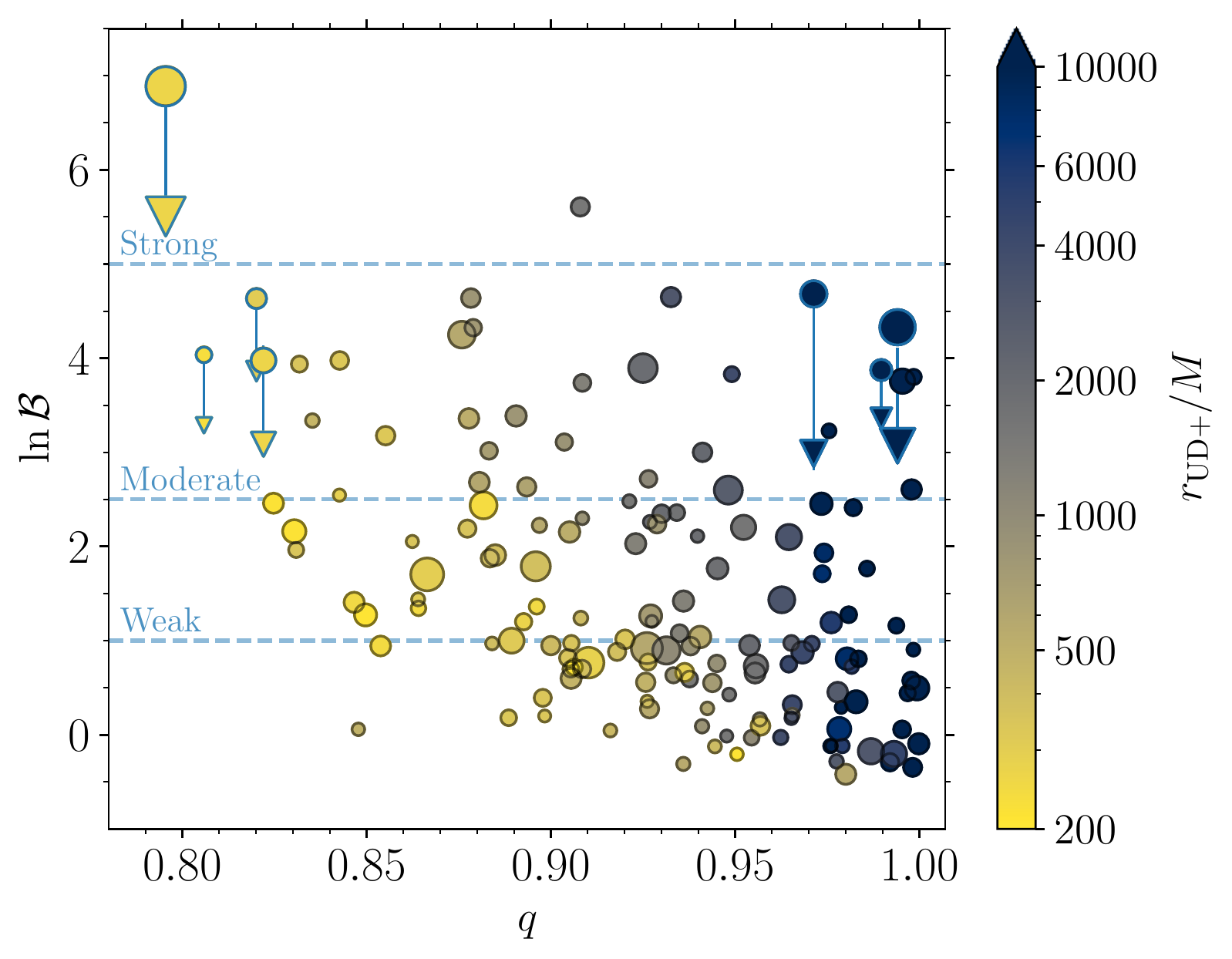}
    \caption{Natural logarithm of the Bayes factor $\mathcal{B}$ as a function of the mass ratio  $q$ for a set of 151 GW signals injected in the endpoint of the up--down instability. 
The critical orbital separation $r_\mathrm{UD +}$ is reported on the color bar and the size of each scatter point is directly proportional to the three-detector SNR.  Horizontal dashed blue lines correspond to the threshold values of the Jeffrey scale for weak, moderate, and strong evidence. The scatter points 
connected by vertical lines are sources that were injected and recovered both with (upper markers, circles) and without (lower markers, triangles)
higher-order modes.
}
\label{fig:BF_q_rplus}
\end{figure*}

We now investigate the distinguishability of up--down sources 
in a wider region of the parameter space.  We construct a set of injections by drawing binaries from the standard uninformative priors;
we sample $q$ and $\chi_{1,2}$ and enforce $\cos\theta_{1,2}$ and $\phi_{12}$ from Eqs.~(\ref{eq:tilt1}--\ref{eq:phi12}). We then impose the following constraints:
\begin{itemize}
\item[(i)] We only consider binaries with $r_\mathrm{UD +}- r_{\rm 20 Hz}>200 M $,
which is a conservative condition to ensure that the analytical instability endpoint well describes binaries that formed in the up--down configuration.
\item[(ii)] 
We further require sources to have $\mathrm{SNR}>20$, thus adopting a threshold that is about twice the current detection limit~\cite{2019PhRvX...9c1040A,2021PhRvX..11b1053A,2021arXiv210801045T,2021arXiv211103606T}.
Spin effects are known to be challenging to measure \cite{2014PhRvL.112y1101V,2016PhRvD..93h4042P,2022PhRvD.106h4040D} and the model-selection problem tackled here inevitably requires loud signals.
\end{itemize}

Our results are shown in Fig.~\ref{fig:BF_q_rplus}, where we report the Bayes factor as a function of the mass ratio $q$, the critical separation $r_{\rm UD+}$, and the SNR.
It is immediate to note that all injections have mass ratios $q\gtrsim 0.8$; this is a direct consequence of
selecting binaries with a large value of
$r_{\rm UD+}\propto (1-q)^{-2} $ 
[cf. Eq.~(\ref{eq:rplus})] and is largely independent of  the total mass $M$ which only enters the source-frame/detector-frame conversion of the  frequency.

Among the 151 sources we select,
we find that 31
present inconclusive evidence in favor of the up--down origin, 45 sources present weak evidence, 73 present moderate evidence, and 2 present strong evidence (recall that we are assuming equal model priors such that the posterior odds and the Bayes factor coincide). 

We find a broad trend indicating that binaries with
more unequal masses
tend to have larger Bayes factors while binaries with close-to-equal masses cover a larger range of
Bayes factors.
The value of $q$ is closely correlated with $r_{\rm UD+}$ from Eq.~(\ref{eq:rplus}), which implies that pinpointing the up--down origin of binaries with lower values of the critical separation $r_{\rm UD+}$ is going to be somewhat easier (as long as $r_{\rm UD+}$ is still sufficiently large that the analytical endpoint provides a reasonable prediction, see above).

Figure~\ref{fig:BF_SNR_realdata} shows Bayes factors  and SNRs for the same set of injections (blue triangles). As expected the two are positively correlated (cf. Fig.~\ref{fig:SeriesSNR}), though with a large dispersion, including several loud sources that still return an inconclusive model selection. 
Even SNRs as large as $\ssim 200$ do not guarantee a decisive model selection
result
since the value of $\mathcal{B}$ strongly depends on the specific parameters of the source.

A key ingredient to this analysis is the inclusion of higher-order emission modes in the adopted waveform model. Higher harmonics can break degeneracies between the mass and spin parameters ~\cite{2019PhRvD.100l3017P,2021PhRvD.103b4042M,2018PhRvD..98h4028C,2021PhRvD.103j4056P}, thus aiding our model selection problem. We further investigate this point by considering seven sources among those with the smaller and larger values of $q$ from our set and repeat their analysis without higher-order modes. As expected, we find that the the Bayes factor decreases,
with differences (in logarithmic scale) that are up to $ \ssim 1.5$.

\subsection{Current gravitational-wave data}
\label{subsec:realdata}

\begin{figure}
    \includegraphics[width=\columnwidth]{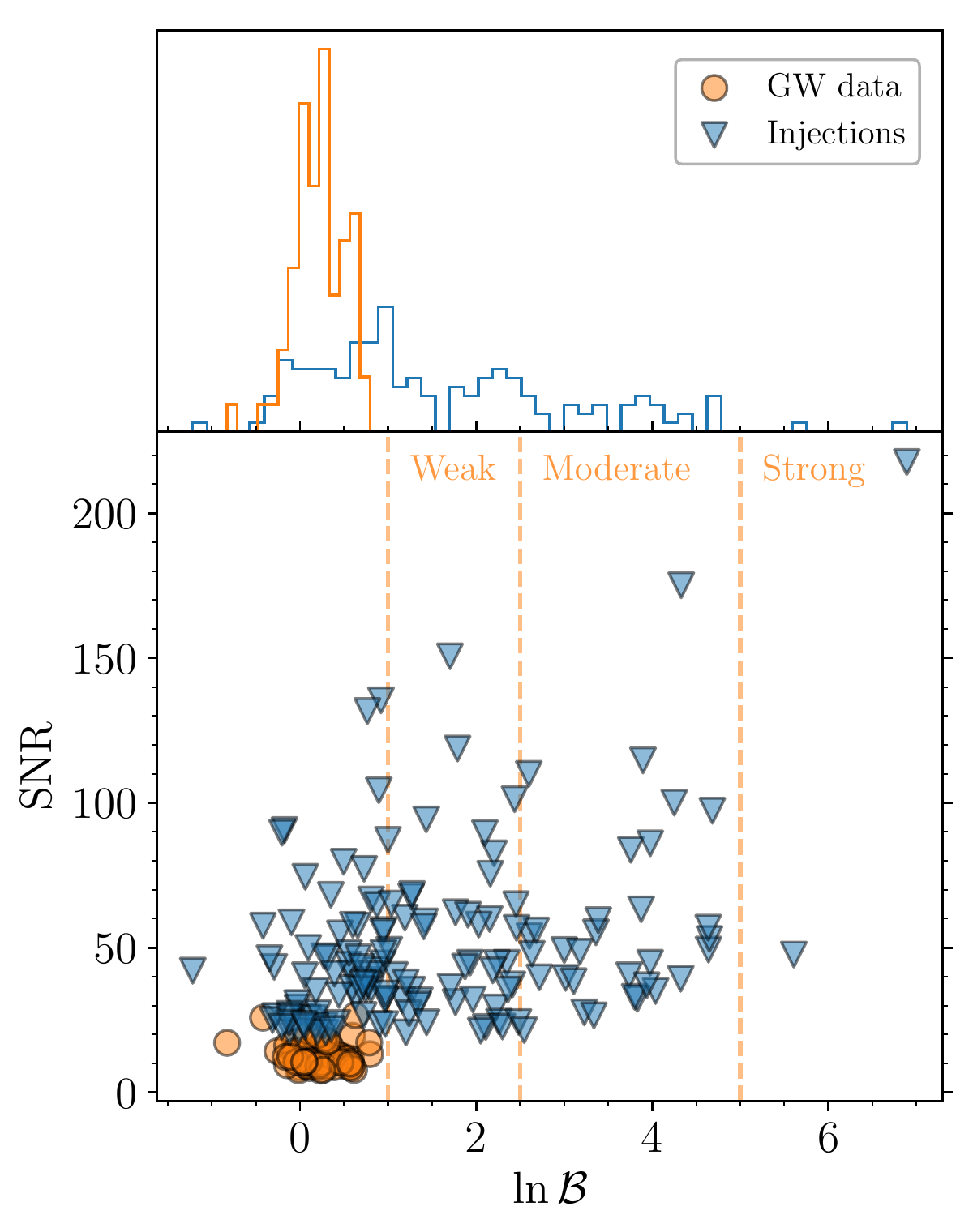}
    \caption{Natural logarithm of the Bayes factor as a function of the SNR for 151 simulated sources (blue triangles) and 69 GW events from GWTC-3 (orange circles). Vertical dashed orange lines indicate the threshold values of the Jeffrey scale for weak, moderate, and strong evidence. The upper panel shows an histogram of the Bayes factors. 
}
\label{fig:BF_SNR_realdata}
\end{figure}

Finally, we apply our model-selection analysis as described in Sec.~\ref{subsec:modelselection} to current GW events reported up to GWTC-3. 
We analyze the 69 binary BH coalescences listed in Table~\ref{tab:GWevents} (see Sec.~\ref{subsec:PE_vanilla}).

Figure~\ref{fig:BF_SNR_realdata} (orange circles) compares 
 the Bayes factor 
 and  the source $\mathrm{SNR}$ 
(estimated using the median of the optimal network
SNR posterior samples).

The Bayes factor in favor of the up--down hypothesis for current GW signals lies within the range $\ln\mathcal{B}\in[-0.8,0.8]$, which is inconclusive. None of the current  events support the up--down endpoint model, but they do not allows us to exclude it either. This is somewhat expected given that SNRs of current event are $\lesssim 30$,
which is unlikely to provide meaningful constraints
(cf. Fig.~\ref{fig:SeriesSNR} and \ref{fig:BF_SNR_realdata}). 
Our finding agrees with previous analyses~\cite{2019PhRvX...9c1040A,2021PhRvX..11b1053A,2021arXiv210801045T,2021arXiv211103606T} indicating that current data provide 
loose constraints on the orientations of individual BH spins, which in turn are key ingredients in the up--down model selection problem.
We conclude that the current catalog of GW events does not contain promising
up--down candidates.

At the same time, we note that the Bayes factor for the entire observed catalog $\sum_i \ln B_i\simeq 15$ shows a preference for the narrow hypothesis $\mathcal{H}_{\rm N}$. Properly quantifying the astrophysical relevance of this finding requires a deeper investigation on the systematics of the single-event $B_i$'s as well as additional population modeling to include selection effects.

\section{Conclusions}
\label{sec:conclusions}

In this paper, we performed parameter estimation of BH binaries that have encountered the up--down instability~\cite{2015PhRvL.115n1102G}.
Binaries that are formed with the spin of the heavier (lighter) BH aligned (anti-aligned) with the orbital angular momentum might enter the LIGO/Virgo band with
significant spin precession.
Their final configuration (i.e., the endpoint of the up--down instability) can be computed
in closed form~\cite{2020PhRvD.101l4037M} and allows us to test the up--down origin of precessing binary BHs.
More ambitiously, one could also target up--down binaries as they become unstable (i.e. $r=r_\mathrm{UD+}$) and start precessing. While worthy of further investigation, the rate of these events is presumably very low.

We presented a statistical approach based on the Savage-Dickey density ratio  for the calculation of the Bayes factor and applied it to both simulated signals (which act as a control set) and current GW events.
%
%
The identification of unstable up--down binaries depends on the source $\mathrm{SNR}$, with higher-order emission modes providing an important contribution. At least within the limited set of injections performed here, we find that SNRs greater than $\ssim 100$ are required. However, this is a necessary but not sufficient condition for the up--down origin to be distinguishable, as the resulting posterior odds
strongly depends on the source parameters. Our model selection analysis is slightly more discriminative 
for sources with
unequal masses and, consequently, with smaller values of $r_\mathrm{UD+}$.
Posterior samples for all the injections presented in this paper are publicly available at  \href{https://github.com/ViolaDeRenzis/updowninjections}{github.com/ViolaDeRenzis/updowninjections}~\cite{datarelease}.

Among the current LIGO/Virgo events,  we do not find promising candidates that could be interpreted as binary systems that were originally aligned in the up--down configuration. This result is not surprising, given the present SNRs which are $\lesssim 30$.

Future LIGO/Virgo upgrades as well as new facilities will largely increase the available statistical sample~\cite{2019PhRvD.100f4060B, 2020LRR....23....3A}. The methodology developed in this paper provides a straightforward, post-processing operation that can be performed on posterior samples from future GW catalogs. Looking ahead, testing the up--down hypothesis is particularly relevant in the context of supermassive BH binaries observed by LISA. Some of those sources are expected to have SNRs as large as $\ssim 3000$~\cite{2017arXiv170200786A} and their spins might be brought to the up--down configuration by interactions with galactic-scale accretion disks~\cite{1975ApJ...195L..65B,2013ApJ...774...43M,2023MNRAS.519.5031S}.

A future detection of the up--down instability presents the opportunity to confirm this prediction
of the general-relativistic two-body problem.

\appendix
\section{Savage-Dickey density ratio}
\label{savdic}
Following the notation introduced in Sec.~\ref{subsec:PE_vanilla}, let us assume that we have some observed data $d$ and two hypotheses such that
\begin{align}
\mathcal{H}_{\mathrm N} : \mathcal{H}_{\mathrm B} \land \gamma=\gamma_{\mathrm{N}}(\varphi)\,.
\end{align}
With this definition,  the evidence of the narrow model is
\begin{align}
\mathcal{Z}(d &| \mathcal{H}_N)=\int  \mathcal{L}(d | \varphi, \mathcal{H}_\mathrm{N})\, \pi(\varphi | \mathcal{H}_\mathrm{N})  \dd\varphi 
\notag \\
 &=\int    \mathcal{L}(d | \varphi, \gamma\!=\!\gamma_{\mathrm{N}}(\varphi),\mathcal{H}_\mathrm{B})\, \pi(\varphi | \gamma\!=\!\gamma_{\mathrm{N}}(\varphi),\mathcal{H}_\mathrm{B}) \dd{\varphi} .
\label{eq:nested_evidence}
\end{align}
One can manipulate the first term in the integrand using  Bayes' theorem,
\begin{align}
  \mathcal{L}(d | \varphi, \gamma\!=\!\gamma_{\mathrm{N}}(\varphi),\mathcal{H}_\mathrm{B})  = \frac{p(\varphi,\gamma\!=\!\gamma_{\mathrm{N}}(\varphi) | d,\mathcal{H}_\mathrm{B})   \mathcal{Z}(d | \mathcal{H}_\mathrm{B}) }{\pi(\varphi,\gamma\!=\!\gamma_{\mathrm{N}}(\varphi) | \mathcal{H}_\mathrm{B})}
\,,
\end{align}
and write the Bayes factor in favor of the narrow model as
%
%
%
%
\begin{align}
\mathcal{B} &=  \frac{\mathcal{Z}(d | \mathcal{H}_\mathrm{N})}{\mathcal{Z}(d | \mathcal{H}_\mathrm{B})}
\notag \\ &=\int \dd{\varphi}p(\varphi,\gamma\!=\!\gamma_{\mathrm{N}}(\varphi) | d,\mathcal{H}_\mathrm{B}) \frac{\pi(\varphi | \gamma\!=\!\gamma_{\mathrm{N}}(\varphi) ,\mathcal{H}_\mathrm{B})}{\pi(\varphi,\gamma\!=\!\gamma_{\mathrm{N}}(\varphi) | \mathcal{H}_\mathrm{B})}.
\end{align}
The rule of conditional probability implies
\begin{align}
\frac{ \pi(\varphi,\gamma\!=\!\gamma_{\mathrm{N}}(\varphi) | \mathcal{H}_\mathrm{B})} { \pi(\varphi | \gamma\!=\!\gamma_{\mathrm{N}}(\varphi) ,\mathcal{H}_\mathrm{B})}  &=  {\pi (\gamma\!=\!\gamma_{\mathrm{N}}(\varphi)| \mathcal{H}_\mathrm{B})}
\notag \\ 
&=  \int  {\pi (\varphi', \gamma\!=\!\gamma_{\mathrm{N}}(\varphi)| \mathcal{H}_\mathrm{B})} \dd \varphi' \,,
\end{align}
where in the second equality we have  explicitly indicated the marginalization over the common parameters. This yields
\begin{equation}
\mathcal{B}=\bigintsss  \frac{\displaystyle  p(\varphi, \gamma\!=\!\gamma_{\rm N}(\varphi) | d, \mathcal{H}_{\rm B}) }{\displaystyle \int  \pi(\varphi',\gamma\!=\!\gamma_{\rm N}(\varphi) | \mathcal{H}_{\rm B}) \dd\varphi'} \; \dd\varphi \,,
\label{sameeqagain}
\end{equation}
which is equal to Eq.~(\ref{eq:BF_SDR0}).
%
%
%
%
%
%

The Savage-Dickey density ratio is recovered by a suitable change of variables:
\begin{align}
\{\varphi,\gamma\}\longrightarrow \{\bar \varphi=\varphi,\bar\gamma\!=\!\gamma-\gamma_{\mathrm{N}}(\varphi)\}\,.
\end{align}
%
The determinant of the resulting Jacobian is
\begin{align}
\det
\begin{pmatrix}
  {\partial \bar\varphi}/{\partial \varphi} & 
    {\partial \bar\varphi}/{\partial \gamma} \\[1ex]
  {\partial \bar\gamma}/{\partial \varphi} & 
     {\partial \bar\gamma}/{\partial \gamma} \\
\end{pmatrix}
=
\det
\begin{pmatrix}
  1 & 
    0 \\[1ex]
 -\dd \gamma_\mathrm{N}/ \dd \varphi& 
      1\\
\end{pmatrix}
=1
\end{align}
such that, for any probability distribution $P$, one can simply write
\begin{align}
P(\varphi,\gamma\!=\!\gamma_{\mathrm{N}}(\varphi))= P(\varphi,\bar \gamma=0)\,.
\end{align}
With this transformation, Eq.~(\ref{sameeqagain}) reduces to
\begin{align}
\mathcal{B}&
=   \frac{\displaystyle \int  p(\varphi, \bar\gamma=0 | d, \mathcal{H}_{\rm B}) \dd\varphi  }{\displaystyle  \int  \pi(\varphi',\bar \gamma= 0 | \mathcal{H}_{\rm B}) \dd\varphi'} 
=  \frac{ p(\bar\gamma=0 | d, \mathcal{H}_{\rm B})}{ \pi(\bar \gamma= 0 | \mathcal{H}_{\rm B})} \,,
\end{align}
as reported in Eq.~(\ref{eq:BF_SDR}), see also Ref.~\cite{2014PhRvD..89j4023C}.


%
%
%
%
%
\vspace{0.7cm}


\acknowledgements
We thank Colm Talbot, Isobel Romero-Shaw, Chris Moore, Francesco Iacovelli, Salvatore Vitale, Neil Cornish,
Sylvia Biscoveanu, Vijay Varma,
and Max Isi  for discussions. V.D.R., D.G., and M.M. are supported by ERC Starting Grant No.~945155--GWmining, Cariplo Foundation Grant No.~2021-0555, MUR PRIN Grant No.~2022-Z9X4XS, and the ICSC National Research Centre funded by NextGenerationEU. D.G. is supported by Leverhulme Trust Grant No.~RPG-2019-350. 
R.B. is supported by Italian Space Agency Grant No.~2017-29-H.0. 
Computational work was performed at CINECA with allocations through INFN, Bicocca, and ISCRA project HP10BEQ9JB.

\vspace{-0.1cm}

\bibliography{updown}

\end{document}